\documentclass[         %
aps,                    
prd,                    
showplace,               
nofootinbib,            
twocolumn,             %
showkeys,               %
preprintnumbers,        %
floatfix]               
{revtex4}               

\usepackage{graphicx,longtable}

\begin{document}

\title{Solutions of Nuclear Pairing}

\author{        A.~B. Balantekin}
\email{         baha@physics.wisc.edu}
\author{Y. Pehlivan}
\email{         yamac@physics.wisc.edu}
\affiliation{  Department of Physics, University of Wisconsin
               Madison, Wisconsin 53706 USA }

\date{\today}
\begin{abstract}
We give the exact solution of orbit dependent nuclear pairing
problem between two non-degenerate energy levels using the Bethe
ansatz technique. Our solution reduces to previously solved cases
in the appropriate limits including Richardson's treatment of
reduced pairing in terms of rational Gaudin algebra operators.
\end{abstract}
\medskip
 \preprint{} \maketitle

Pair correlations are manifest in a remarkable range of
quantum many-body systems. Originally the idea of pairing and the
methods were developed in the context of superconductivity in
macroscopic systems by Bardeen, Cooper and Schrieffer
\cite{Bardeen:1957kj} and by Bogoliubov \cite{Bogoliubov}. In
nuclear physics, it has been long known that the independent
particle picture must be improved with the addition of a
nondiagonal two-particle force as evidenced by the absence of
single particle excitations at low energies \cite{Mottelson} and
soon the idea of the pairing was carried over \cite{BMP,Belyaev}.
But methods borrowed from
superconductivity in infinite systems were inconvenient for the
finite nucleus because the former is based on wave functions with
indefinite number of particles. Although it is known that
the pairing effects play a major role in determining nearly all
nuclear properties including the excitation spectra, the
transition probabilities and the equilibrium shape,
an exact number-conserving solution to nuclear pairing problem is
still lacking except in three limits. The limit in which
single particle energy levels are degenerate and all pairing strengths
are the same was solved by Kerman using the
quasi-spin algebra \cite{kerman1}. Later Richardson solved the
limit in which the single particle energy levels are nondegenerate
but the pairing strengths are the same \cite{rich}.
Finally, the limit in which the single particle levels have
different pairing strengths but are all degenerate is solved
by Pan \textit{et al} \cite{Pan:1997rw} and by Balantekin
\textit{et al} \cite{Balantekin:2007qr,Balantekin:2007vs}. On the
other hand, in many cases it is the interplay between the one-body
and the two-body effects which determine the equilibrium
properties of the nucleus and a full solution of the nuclear
pairing problem is highly desirable as a realistic model. The
purpose of this Letter is to present an exact solution of the
pairing problem away from the above mentioned limits, i.e., with
orbit dependent pairing strengths and non-degenerate single
particle energies, in the presence of two nuclear levels.

The problem is described by the Hamiltonian
\begin{equation}
\label{1} \hat{H}_P=\sum_{jm} \epsilon_j a^\dagger_{j\>m} a_{j\>m}
- |G|\sum_{jj'}c^*_jc_{j'} \hat{S}^+_j \hat{S}^-_{j'}.
\end{equation}
Here $c_j$'s are the occupation amplitudes and the quasi-spin
operators $S^{\pm,0}_j$ are given by
\begin{eqnarray}\label{2}
\hat{S}^+_j&=&\left(\hat{S}^-_j\right)^\dagger=\sum_{m>0}
(-1)^{(j-m)} a^\dagger_{j\>m}a^\dagger_{j\>-m},
\nonumber \\
\hat{S}^0_j&=&\sum_{m>0}\frac{1}{2}
\left(a^\dagger_{j\>m}a_{j\>m}+a^\dagger_{j\>-m}a_{j\>-m}-1
\right).
\end{eqnarray}
The interaction described by the Hamiltonian (\ref{1}) is known as
separable pairing, since the interaction strength is the
multiplication of two numbers (i.e. $c_jc_{j^\prime}$).
\footnote{Certain nonseparable pairing models are also known to be
integrable such as the rational, trigonometric and hyperbolic
Gaudin magnet Hamiltonians \cite{gaudin,Dukelsky:2001fe}. These
three sets of Hamiltonians are mutually commuting and
simultaneously diagonalizable. They can be combined in various
ways to built other integrable pairing models. The separable
Hamiltonian in (\ref{1}), however, does not commute with any
of the Gaudin magnet Hamiltonians and hence do not belong to the
class of pairing problems which can be approached in this manner
except when all $c_j$'s are the same.}

The limit of the pairing problem described by (\ref{1}) in which
all single particle energy levels are degenerate (leading to the
first term being a constant for a given number of pairs) and all
$c_j$'s are the same can be treated using the quasi-spin algebra
generated by $S^{\pm,0}_j$ \cite{kerman1}.

The case when all $c_j$'s are still the same, but single particle
energies are non-degenerate was treated by Richardson \cite{rich}
who obtained energy eigenvalues and eigenstates in terms of the
rational Gaudin algebra operators
\begin{equation}\label{3}
\tilde{J}^+(x)=\sum_j\frac{1}{2\varepsilon_j-x}S_j^+,
\end{equation}
dependent on solutions of certain Bethe ansatz equations. Here
$\varepsilon_j=\epsilon_j/|G|$ are the scaled single particle
energies. More recently, the authors of Ref. \cite{Pan:1997rw}
solved the case when the single particle energies are degenerate,
occupation amplitudes of individual orbits ($c_j$'s) are
different, but the shell is at most half full, using the step
operators
\begin{equation}\label{4}
S^+(x)=\sum_j\frac{c_j^*}{1-|c_j|^2x}S_j^+.
\end{equation}
This solution was generalized to the case when the shell is more
than half full in Refs. \cite{Balantekin:2007vs} and
\cite{Balantekin:2007qr}.

To give an
exact solution for the most general problem
with two orbits we use the step
operators
\begin{equation}\label{5}
J^+(x)=\sum_j\frac{c_j^*}{2\varepsilon_j-|c_j|^2x}S_j^+
\end{equation}
which approach to those in given (\ref{3}) and (\ref{4}) in
the appropriate limits with a rescaling of the variable $x$.

To present our solution, we write (\ref{1}) as
\begin{equation}
\label{6} \hat{H}=\frac{\hat{H}_P}{|G|}= \sum_{j} 2\varepsilon_j
\hat{S}_j^0 - \sum_{jj'}c^*_jc_{j'} \hat{S}^+_j
\hat{S}^-_{j'}+\sum_j\varepsilon_j\Omega_j,
\end{equation}
where we used (\ref{2}) and divided by $|G|$
to work with scaled single particle energies
$\varepsilon_j=\epsilon_j/|G|$. In the last term of Hamiltonian
(\ref{6}), $\Omega_j$ denotes the maximum number of pairs that can
occupy level $j$. Although this term is a constant, we keep it in
order to guarantee that the energy of the empty shell is zero. The
empty shell is represented by the lowest weight state of all the
quasispin operators defined in (\ref{2}) and we denote it by
$|0\rangle$.

To find the energy eigenstates with one pair, we form the
Bethe ansatz state $J^+(x)|0\rangle$ where $J^+(x)$ is given by
(\ref{5}). Using the quasi-spin algebra generated by the
operators introduced in (\ref{2}), one can show that the state
$J^+(x)|0\rangle$ is an eigenstate of Hamiltonian (\ref{6}) with
the energy
\begin{equation}\label{10}
E_1=-\frac{\alpha x}{\beta-x}
\end{equation}
if $x$ is chosen so that the Bethe ansatz equation
\begin{equation}\label{11}
\sum_{j}\frac{\Omega_j|c_j|^2}{2\varepsilon_j-|c_j|^2x}
=\frac{\beta}{\beta-x}
\end{equation}
is satisfied. Here $\alpha$ and $\beta$ are given by
\begin{equation}\label{9}
\alpha
=2\frac{\varepsilon_{j_2}|c_{j_1}|^2-\varepsilon_{j_1}|c_{j_2}|^2}
{|c_{j_1}|^2-|c_{j_2}|^2} \quad \quad
\beta=2\frac{\varepsilon_{j_1}-\varepsilon_{j_2}}{|c_{j_1}|^2-|c_{j_2}|^2}.
\end{equation}
This result generalizes to the state
\begin{equation}\label{12}
J^+(x_1)J^+(x_2)\dots J^+(x_N)|0\rangle,
\end{equation}
with $N$ pairs where all the variables $x_k$ are different from
one another. Here we assume that the shell is at most half full,
i.e., $N\leq N_{max}/2$ where $N_{max}$ is the maximum number of
pairs that can occupy the shell. As will be described below, if
the shell is more than half full, it is easier to work with the
hole pairs instead of the particle pairs. One can show that the
state (\ref{12}) is an eigenstate if the parameters $x_k$ satisfy
the Bethe ansatz equations
\begin{equation}\label{13}
\sum_{j}\frac{\Omega_j|c_j|^2}{2\varepsilon_j-|c_{j}|^2x_k}
=\frac{\beta}{\beta-x_k} +\sum_{n=1(\neq k)}^N\frac{2}{x_n-x_k}.
\end{equation}
If these equations are satisfied, then the energy of the state
(\ref{12}) is given by
\begin{equation}\label{14}
E_N=-\sum_{n=1}^N\frac{\alpha x_n}{\beta-x_n}.
\end{equation}
In general, equations of Bethe ansatz have more than one
solutions. Each set of numbers $(x_1,x_2,\dots,x_N)$ satisfying
the Bethe ansatz equations gives us an eigenstate in the form of
(\ref{12}). Solutions may involve complex numbers but since
they always come in complex conjugate pairs, the energy given in
(\ref{14}) is always real.

It is easy to see that the two particular cases, namely the orbit
independent (i.e., reduced) coupling case solved by Richardson and
the degenerate single particle energy case solved by Pan
\textit{et al} and Balantekin \textit{et al} can be obtained from
(\ref{13}) and (\ref{14}).

\textit{Richardson Limit} ($\beta\to\infty$): In the limit where
the occupation amplitudes are all equal to each other (i.e.
$c_j=c$), the step operators of (\ref{5}) approach to rational
Gaudin algebra operators given in (\ref{3}) with a rescaling
of the variable $x$ as $x'=|c|^2x$. Consequently, the eigenstates
(\ref{12}) approach to those proposed by Richardson \cite{rich}.
Since we have $\beta\to\infty$ and $\alpha/\beta\to |c|^2$ in this
limit, Bethe ansatz equations (\ref{13}) and the energy (\ref{14})
become
\begin{equation}\label{15}
\sum_{j}\frac{\Omega_j}{2\varepsilon_j-x'_k}=\frac{1}{|c|^2}
+\sum_{n=1(\neq k)}^N\frac{2}{x'_n-x'_k},
\end{equation}
\begin{equation}\label{16}
E_{N}=\sum_{n=1}^{N}x'_{n}.
\end{equation}
These equations are those obtained by Richardson in \cite{rich}.

\textit{Degenerate Limit}($\beta\to 0$): In the limit where single
particle energies are degenerate (i.e.
$\varepsilon_j=\varepsilon$), the step operators given in
(\ref{5}) approach to those in (\ref{4}) with a
rescaling of the variable $x$ as $x''=x/2\varepsilon$. Therefore
the eigenstates (\ref{12}) approach to those found in Refs.
\cite{Pan:1997rw,Balantekin:2007qr,Balantekin:2007vs}. Since
$\beta\to 0$ in this limit, one should distinguish between the
case in which all variables $x_k$ are different from zero and the
case in which one of the variables is zero. Consequently, we
obtain two types of eigenstates in this limit.

From (\ref{13}), it can be easily seen that if all $x_k$ are
different from zero, then the Bethe ansatz of (\ref{13})
becomes
\begin{equation}\label{17}
\sum_j\frac{\Omega_j|c_j|^2}{1-|c_j|^2x_k'' } %
=\sum_{n=1(\neq k)}^N \frac{2}{x_n''-x_k''}
\end{equation}
in the degenerate limit. Corresponding energy given in
(\ref{14}) approaches to a constant value given by\footnote{These
states are called ``zero energy states''  in Refs.
\cite{Pan:1997rw,Balantekin:2007qr,Balantekin:2007vs} since the
degenerate single particle energy $\varepsilon$ is taken to be
zero in which case the energy (\ref{18}) also vanishes.}
\begin{equation}\label{18}
E_N=2N\varepsilon.
\end{equation}

\begin{figure}
  \includegraphics
  {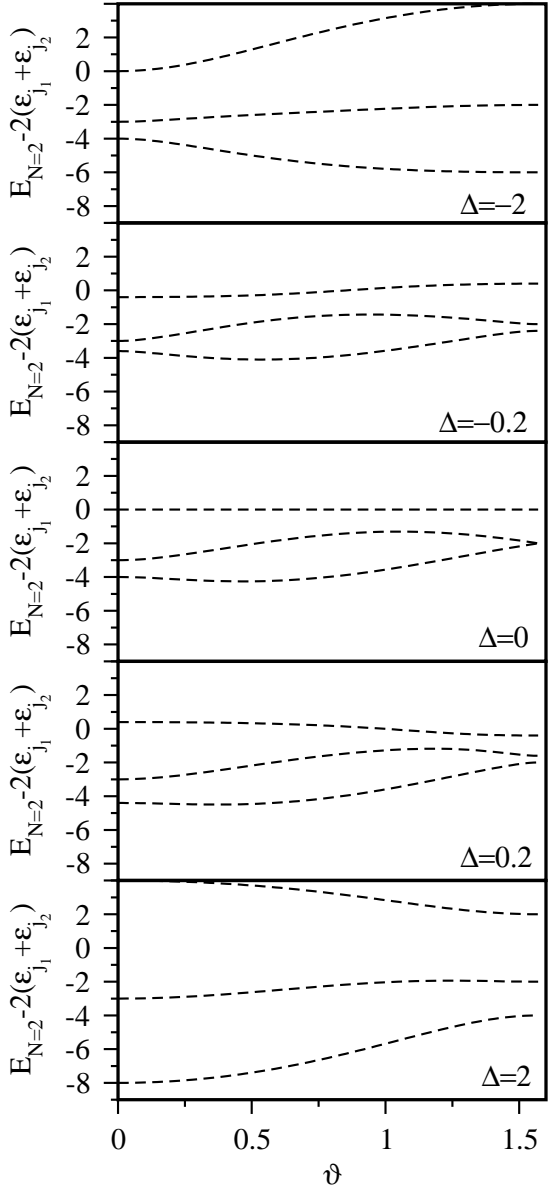}\\
\caption{Energies for two particle pairs with $j_1=3/2$ and
$j_2=5/2$. Here $\vartheta$ describes the occupation amplitudes as
described in the text and
$\Delta=\varepsilon_{j_1}-\varepsilon_{j_2}$ is the separation
between single particle levels.} \label{graph}
\end{figure}

Some solutions of the equations of Bethe ansatz (\ref{13}) involve
one variable, say $x_1$, which approaches to zero as the single
particle levels become degenerate. Note that there can be at most
one such variable since all $x_k$ must be different from one
another. In this case, the limit of the Bethe ansatz equations
(\ref{13}) for the rest of the variables (i.e. for $k\geq 2$) when
$\beta\to 0$ is
\begin{equation}\label{19}
\sum_j\frac{\Omega_j|c_j|^2}{1-|c_j|^2x_k''}
=-\frac{2}{x_k''}+\sum_{n=2(n\neq k)}^N\frac{2}{x_n'' -x_k''}.
\end{equation}
Using (\ref{14}) and the limit of
(\ref{13}) for $k=1$ as $\beta\to 0$, the energy of the
corresponding state can be found as
\begin{equation}\label{20}
E_N=2N\varepsilon-\sum_j\Omega_j|c_j|^2
+\sum_{n=2}^{N}\frac{2}{x_n''}.
\end{equation}
Equations (\ref{17}-\ref{20}) are those found in Refs.
\cite{Pan:1997rw,Balantekin:2007qr,Balantekin:2007vs}.

If the shell is more than half full, it is easier to work with
hole pairs instead of particle pairs. Fully occupied shell is
described by the highest weight state of all the quasispin
operators given in (\ref{2}). We denote this state by
$|\bar{0}\rangle$. The energy of the fully occupied shell is
\begin{equation}\label{21}
E_{N_{max}}=\sum_j\left(2\varepsilon_j-|c_j|^2\right)\Omega_j.
\end{equation}
Here $N_{max}$ denotes the maximum occupancy number of the shell.
We can create hole pairs (or, equivalently, annihilate particle
pairs) by acting on the fully occupied shell with the operators
\begin{equation}\label{22}
J^-(y)=\sum_j\frac{c_j^*}{2\varepsilon_j-|c_j|^2y}S_j^-.
\end{equation}
One can show that the state $J^-(y)|\bar{0}\rangle$ which has one
hole pairs ($N_{max}-1$ particle pairs), is an eigenstate of the
Hamiltonian with the energy
\begin{equation}\label{24}
E_{N_{\max }-1}=E_{N_{\max }}+\alpha \frac{y-2}{\beta-y},
\end{equation}
if $y$ obeys the Bethe ansatz equation
\begin{equation}\label{25}
\sum_{j}\frac{\Omega_j|c_j|^2}{2\varepsilon_j-|c_j|^2y}
=\frac{2-\beta}{\beta-y}.
\end{equation}
Similarly, for $\bar{N}$ hole pairs ($N_{max}-\bar{N}$ particle
pairs), we write the Bethe ansatz state
\begin{equation}\label{26} J^-(y_1)J^-(y_2)\dots
J^-(y_{\bar{N}})|\bar{0}\rangle.
\end{equation}
Here $\bar{N}<N_{max}/2$ and all the variables $y_k$ are different
from each other. This state is an eigenstate of the Hamiltonian
(\ref{6}) with the energy
\begin{equation}\label{27}
E_{N_{\max }-\bar{N}}=E_{N_{\max }}+\alpha
\sum_{n=1}^{\bar{N}}\frac{y_n-2}{\beta-y_n}
\end{equation}
if the variables $y_k$ obey the equations of Bethe ansatz given by
\begin{equation}\label{28}
\sum_j\frac{\Omega_j|c_j|^2}{2\varepsilon_j-|c_j|^2y_k}
=\frac{2-\beta}{\beta-y_k} +\sum_{n=1(\neq
k)}^{\bar{N}}\frac{2}{y_n-y_k}.
\end{equation}
In the limit where the single particle energies are degenerate,
(\ref{24})-(\ref{28}) approach to those found in Ref.
\cite{Balantekin:2007vs}.

Next, we provide exact solutions of the equations of Bethe ansatz
for $N=1$ and $N=2$ pairs to illustrate the technique.
Since the occupancy amplitudes are normalized as
$|c_{j_1}|^2+|c_{j_2}|^2=1$ we can set
$c_{j_1}=\sin\vartheta$ and $c_{j_2}=\cos\vartheta$
where $0\leq \vartheta \leq \pi$. For one pair, one has to solve
the Bethe ansatz equation given in (\ref{11}). This equation
has two distinct solutions which lead to two different energy
eigenvalues when substituted in (\ref{10}):
\begin{widetext}
\begin{equation}\label{30}
E_{N=1}=\frac{\Omega_{j_1}\sin ^{2}\vartheta +\Omega_{j_2}\cos
^{2}\vartheta }{2}-\left( \varepsilon _{j_1}+\varepsilon
_{j_2}\right) \pm
\frac{1}{2}\sqrt{4\Delta^{2}-4\left(\Omega_{j_1}\sin ^{2}\vartheta
-\Omega_{j_2}\cos ^{2}\vartheta \right) \Delta + \left(
\Omega_{j_1}\sin ^{2}\vartheta +\Omega_{j_2}\cos ^{2}\vartheta
\right)^{2}}
\end{equation}
\end{widetext}
where we defined $\Delta=\varepsilon_{j_1}-\varepsilon_{j_2}$.

For $N=2$, one should solve the equations of Bethe ansatz
(\ref{13}) for $x_1$ and $x_2$. For two pairs, there are three
distinct solutions of the equations of Bethe ansatz leading to
three energy eigenvalues. Since the resulting analytical
expressions for the energy eigenvalues are long, we find it more
convenient to present them in an alternative form. Starting from
the Bethe ansatz equations (\ref{13}) and the energy (\ref{14}),
it can be shown that for $N=2$ the three energy eigenvalues of the
Hamiltonian are given by
\begin{equation}\label{31}
E_{N=2}=\frac{w}{\cos(2\vartheta)}+2\left(\varepsilon_{j_1}+\varepsilon_{j_2}\right)
\end{equation}
where $w$ is one of the three roots of the following cubic
polynomial:
\begin{widetext}
\begin{eqnarray}\label{32}
&&2w^{3}+2\cos2\vartheta\left(3P-2\right) w^{2}+\cos^{2}2\vartheta
\left[4P\left(P-1\right)+8\Delta \left( \Omega _{j_1}\sin
^{2}\vartheta -\Omega _{j_2} \cos ^{2}\vartheta \right) -8\left(
\Delta +\sin ^{2}\vartheta \right) \left( \Delta -\cos
^{2}\vartheta \right)\right. \nonumber \\  && \left. -\sin
^{2}2\vartheta \left( \Omega _{j_1}+\Omega_{j_2}\right) \right] w
-8\Delta \cos ^{3}2\vartheta \left[\Omega _{j_2}\left( \Omega
_{j_2}-1\right) \cos ^{4}\vartheta -\Omega _{j_1}\left( \Omega
_{j_1}-1\right) \sin ^{4}\vartheta +\Delta P\right] =0.
\end{eqnarray}
\end{widetext}
Here we defined $P=\Omega_{j_1}\sin^{2}\vartheta+\Omega_{j_2}\cos
^{2}\vartheta$. Roots of this polynomial can be found analytically
for any set of parameters. For example, in Figure \ref{graph}, we
plot the exact energy eigenvalues for two pairs found from
(\ref{31}) and (\ref{32}) for $j_1=3/2$ and $j_2=5/2$. We show how
the energy eigenvalues change with the occupation amplitudes for
different values of $\Delta=\varepsilon_{j_1}-\varepsilon_{j_2}$.
In the degenerate case (i.e. when $\Delta=0$), one of the energy
eigenvalues is constant in agreement with (\ref{18}). When we
slowly deviate from the degenerate case, for example when
$\Delta=\pm 0.2$, this energy eigenvalue is still independent of
the occupation amplitudes for the most part.

In this Letter, we presented the exact solution of the orbit
dependent pairing problem between two nondegenerate energy levels.
We showed that this solution reproduces the previously known
results in the reduced pairing limit and the degenerate limit
\cite{rich,Pan:1997rw,Balantekin:2007vs,Balantekin:2007qr}. Since
the solution is exact, it is also valid away from these limits and
it can be used to study the interplay between the one-body effects
and the orbit dependent pairing effects in a realistic model. We
also believe that the technique presented here is an important
step towards the exact analytical solution of the most general
pairing problem.

 This  work   was supported in  part  by   the
NSF Grant No. PHY-0555231, and  in  part by  the University of
Wisconsin Research Committee   with  funds  granted by the
Wisconsin Alumni  Research Foundation.

\end{document}